\documentclass[aps,preprint]{revtex4}%
\usepackage{amsfonts}
\usepackage{amsmath}
\usepackage{amssymb}
\usepackage{graphicx}%
\begin{document}
\title{A numerical study of the correspondence between paths in a causal set and
geodesics in the continuum}
\author{Raluca Ilie\footnote{Present address: Department of Atmospheric, Oceanic, and Space Sciences,
University of Michigan, Ann Arbor, Michigan 48109 USA.}, Gregory
B. Thompson\footnote{Present address: Department of Physics and
Astronomy, University of Toledo, Toledo, Ohio 43606 USA}, and
David D. Reid\footnote{Present address: Department of Physics,
University of Chicago, 5720 S. Ellis Ave., Chicago, IL 60637 USA;
contact person (dreid@uchicago.edu).} } \affiliation{Department of
Physics and Astronomy, Eastern Michigan University, Ypsilanti, MI
48197 USA} 

\begin{abstract}
This paper presents the results of a computational study related
to the path-geodesic correspondence in causal sets. For intervals
in flat spacetimes, and in selected curved spacetimes, we present
evidence that the longest maximal chains (the longest paths) in
the corresponding causal set intervals statistically approach the
geodesic for that interval in the appropriate continuum limit.

\end{abstract}
\startpage{1}
\endpage{ }
\maketitle

\section{Introduction}

The causal set program proposes one of a number of approaches to the problem
of quantum gravity \cite{prl}-\cite{rdsrev}. The key aspect of this approach
is that it postulates that spacetime is discrete rather than continuous. A
causal set is a set $C$ of elements $x_{i}\in C,$ and an order relation
$\prec$, such that the set $C=\{x_{i},\prec\}$ obeys properties which make it
a good discrete counterpart for continuum spacetime. These properties are that
(a) the set is transitive: $x_{i}\prec x_{j}\prec x_{k}\Rightarrow x_{i}\prec
x_{k}$; (b) it is noncircular, $x_{i}\prec x_{j}$ and $x_{j}\prec
x_{i}\Rightarrow x_{i}=x_{j}$; (c) it is locally finite in the sense that the
number of elements between any two related elements $x_{i}\prec x_{j}$ is
finite, i.e., $\left\vert [x_{i},x_{j}]\right\vert <\infty$; and (d) it is
reflexive, $x_{i}\prec x_{i}$ $\forall$ $x\in C$. The action of the order
relation is to mimic the causal ordering of events in macroscopic spacetime.
Since all events in spacetime are not causally related, then not all pairs of
elements in the set are ordered by the order relation. Hence, a causal set is
a partially ordered set.

The continuing development of the causal set program can be roughly divided
into two categories: kinematics and dynamics. Development of a causal set
dynamics is the ultimate goal of this program and would represent its version
of a theory of quantum gravity. However, the ability to investigate the
implications of this dynamics will require development of the mathematical and
computational tools needed to describe the physics of causal sets. This
collection of techniques make up causal set kinematics. Of particular interest
here is the correspondence between spacetime as a causal set and macroscopic
spacetime. If a causal set comprises the true structure of spacetime it must
produce a four-dimensional Lorentzian manifold in macroscopic limits.

The goal of determining how we can recognize and extract
manifold-like properties from a partially ordered set is a
kinematic issue of significant current interest. This paper
reports on a numerical study concerning one aspect of this goal;
specifically, we address the correspondence between paths in a
causal set and geodesics in continuum spacetimes. To the best of
our knowledge, this is the first published investigation of this
correspondence that includes curved spacetimes. In what follows we
first describe the key ideas related to this correspondence, we
then describe the details of how our calculations were performed,
and finally we present our results and conclusions.

\section{The Path-Geodesic Correspondence}

The properties of causal sets stated previously do not guarantee that the sets
will display the manifold-like behavior required by the physical theory of
causal sets we ultimately seek. In fact it is well known that most sets
satisfying those properties will not be consistent with points in a spacetime
manifold. A necessary (but not sufficient) requirement for a causal set to be
like a manifold is that it can be embedded into a manifold uniformly with
respect to the metric. An embedding of a causal set is a mapping of the set
onto points in a Lorentzian manifold in such a way that the lightcone
structure of the manifold preserves the ordering of the set. The most probable
embeddings will be uniform if the mapping corresponds to selecting points in
the manifold via a Poisson process. When this latter requirement is met, the
embedding is said to be \textit{faithful}.

Recall that the length of the geodesic between two causally related events
gives the longest proper time between those events. To see what the most
natural analog to geodesic length is for causal sets we must first define a
few terms. A \textit{link,} $\preceq$, in a causal set is an irreducible
relation; so, $x_{i}\preceq x_{k}$ iff $\nexists$ $x_{j}\ni x_{i}\prec
x_{j}\prec x_{k}$. A \textit{chain} in a causal set is a set of elements for
which each pair is related; for example, $x_{a}\prec x_{b}\prec\cdots\prec
x_{z-1}\prec x_{z}$ is a chain from $x_{a}$ to $x_{z}$. A \textit{maximal
chain}, or \textit{path}, is a chain consisting only of links, such as
$x_{a}\preceq x_{b}\preceq\cdots\preceq x_{z-1}\preceq x_{z}$.

Myrheim argued \cite{myrheim} that the length of the longest path between two
related elements in a causal set is the most natural analog for the geodesic
length between two causally connected events in spacetime. The length of a
path is defined to be the number of links in that maximal chain. A more formal
statement, that we refer to as the \textit{Myrheim length conjecture} is the following:

\begin{quote}
Let $\{C_{n}\}$ be a sequence of causal sets $C_{n}=\{x_{i},\prec\}$ of
increasing density $\rho$ that are faithfully embeddable into a Lorentzian
manifold $\mathcal{M}$ by a map $g:C_{n}\rightarrow\mathcal{M}$. Then, in the
limit $\rho\rightarrow\infty$, the expected length of the longest maximal
chains between ordered pairs $(x_{i},x_{j})\in C_{n}$ is directly proportional
to the geodesic lengths between their images $g(x_{i},x_{j})\in\mathcal{M}$.
\end{quote}

Brightwell and coworkers have proven the above statement for the case when the
Lorentzian manifold is Minkowski space \cite{bng}. Their result uses the fact
that, in $d$-dimensional Minkowski space, the volume $V$ of a region of
spacetime bounded by the lightcones of two causally connected events is
$V=m_{d}\tau^{d}$, where $\tau$ is the longest proper time between those
events and $m_{d}$ is a constant. In curved spacetimes (with nonconstant
curvature) that relationship breaks down in that (a) any proportionality
factor between $V$ and $\tau^{d}$ will be a function of spacetime and (b) the
dependence of $V$ on $\tau$ may include additional powers of $\tau$. While it
is widely believed that the Myrheim length conjecture is true for curved
spacetimes, no proof has been achieved.

Underlying the development of the causal set program is a much wider
conjecture, termed the \textit{Hauptvermutung}, which effectively says that in
the appropriate limits, a physically interesting causal set will bring forth
all of the structure of an approximately unique spacetime manifold
\cite{rdsrev, lb-phd}. This conjecture might be stated the following way:

\begin{quote}
Let there be a random process that produces a faithfully embeddable causal set
$C$. Then, if $\exists$ maps $g_{1}$ and $g_{2}$ $\ni g_{1}:C\rightarrow
\mathcal{M}_{1}$ and $g_{2}:C\rightarrow\mathcal{M}_{2}$, then $\mathcal{M}%
_{1}$ and $\mathcal{M}_{2}$ are approximately isometric.
\end{quote}

\noindent One only requires an approximate isometry to allow for random
fluctuations where \textquotedblleft approximately isometric\textquotedblright%
\ might mean that the manifolds are close in the sense discussed in
\cite{lb-close}. This is the main conjecture of causal set kinematics to which
the previously mentioned length conjecture is really just a corollary. Some
evidence for the validity of the Hauptvermutung has already been established
\cite{lb-dm} but a full proof has not yet been formulated. However, if the
correspondence between a causal set and a manifold is to be (nearly) unique,
then not only should the length of the longest paths produce the geodesic
length, but the longest paths themselves should produce the geodesic.

In this work we look for evidence that the longest paths in faithfully
embeddable causal sets approach the appropriate geodesic curve as the density
of causal set elements increases. This work also constitutes an indirect test
of the Myrheim length conjecture in curved spacetimes, because we use this
conjecture to identify which paths might produce the geodesic upon embedding.
The details of how this is done are given below.

\section{Computational Details}

\subsection{Overall Approach}

If we consider points $y_{i}$ in a spacetime manifold $\mathcal{M}$, we can
define an interval $I_{\mathcal{M}}$ between two causally related points
$y_{\text{p}}$ and $y_{\text{f}}$ as the intersection of the future of the
pastmost point $y_{\text{p}}$ with the past of futuremost point $y_{\text{f}}%
$. Let us also define $S_{\mathcal{M}}$ as the geodesic between $y_{\text{p}}$
and $y_{\text{f}}$. Then, our general approach is as follows: (a) we generate
many faithfully embeddable causal sets $C=\{x_{i},\prec\}$ containing $N$
elements in the form of intervals $I_{C}(x_{1},x_{N})$ by selecting points in
$I_{\mathcal{M}}(y_{\text{p}},y_{\text{f}})$ via a Poisson process; (b) in
each $I_{C}$ we sample the longest paths $S_{C}$ between $x_{1}$ and $x_{N}$
and determine the average goodness-of-fit between the images of the $S_{C}$ in
$I_{\mathcal{M}}$ and $S_{\mathcal{M}}$; (c) we then repeat steps (a) and (b)
for larger causal sets and examine the behavior of the average goodness-of-fit
as $N$ increases.

\subsection{Random Sprinklings}

As stated previously, we generate faithfully embeddable causal sets by
selecting points from a manifold via a Poisson process -- this is also called
a random sprinkling into the manifold. Justification for why this should be a
Poisson process is given elsewhere \cite{dm-phd}-\cite{ddr-prd}. The procedure
by which we perform these sprinklings is described in \cite{ddr-prd}; but we
briefly explain it here for completeness.

The spacetimes used in this study are given by the metric
\begin{equation}
ds^{2}=\Omega^{2}\eta_{\alpha\beta}dx^{\alpha}dx^{\beta},
\label{conformal metric}%
\end{equation}
where $\Omega^{2}$ is the conformal factor and $\eta_{\alpha\beta}$ is the
Minkowski tensor. The sprinklings are performed by a (double) rejection method
similiar to the methods described in \cite{dm-phd} and \cite{numrec}. The
interval is enclosed in a box and points are randomly selected within this
box; if a selected point is outside the interval it is rejected, otherwise, it
is kept -- this is the first rejection. In flat spacetime $(\Omega^{2}=1)$
this first rejection provides the desired uniform distribution of points. In
curved spacetimes, however, remaining points face a second rejection to ensure
that the final points are distributed uniformly with respect to the volume
form $\Omega^{d}$, where $d$ is the dimension of the spacetime. Each remaining
point is associated with a random number $w$ within the range $0<w<\Omega
_{\max}^{d}$, where $\Omega_{\max}^{d}$ is the maximum value of the volume
form within $I_{\mathcal{M}}$. If $w$ is greater than the value of the volume
form evaluated at the point in question, the point is rejected; otherwise, it
is kept. This process continues until $N$ points are sprinkled into
$I_{\mathcal{M}}$.

When partially ordered according to their causal relations in $I_{\mathcal{M}%
}$, these $N$ points, without the manifold, comprise a faithfally embeddable
causal set interval $I_{C}$. At the end of the sprinkling process, the
selected points are sorted according to increasing value of the time
coordinate. The corresponding elements in $I_{C}$ are labeled with a number
from 1 to $N$. However, keep in mind that, exclusive of $x_{1}$ and $x_{N}$,
this \textit{natural labeling} does not necessarily correspond to an ordering
of the elements. For example, element $x_{25}$ may preceed element $x_{50}$
($x_{25}\prec x_{50}$) or they may not be related in any way.

\subsection{Finding the Paths}

Having formed the causal set interval, the longest paths between $x_{1}$ and
$x_{N}$ must be found. Starting with element $x_{1}$ we form the set of
elements $D_{1}$ linked to its \textquotedblleft future\textquotedblright;
that is, $D_{1}=\{x_{j}\in I_{C}|x_{1}\preceq x_{j}\}$. For each of these
\textit{daughters} of $x_{1}$ we calculate the length $L(x_{j})$ of the path
from $x_{j}$ to $x_{N}$ using an algorithm due to Sorkin \cite{lisp}. We save
only those elements of $D_{1}$ for which $L(x_{j})$ is maximum. This forms the
subset $Y_{D1}$ of the \textit{youngest} daughters of $x_{1}$. Every element
in $Y_{D1}$ is in a longest path from $x_{1}$ to $x_{N}$. Then, to complete a
longest path we can choose any element of $Y_{D1}$ and repeat the process
until we reach an element for which the only daughter is $x_{N}$.

In general, there are many paths that traverse an interval. In fact, the
number of paths can become quite large. Table 1 lists the average number of
longest paths (for 10 trials) for various interval sizes $N$ for causal sets
generated by sprinkling into 1+1-dimensional Minkowski space. As the table
shows, the number of longest paths is on the order of 10$^{10}$ for intervals
as small as $N=2000$. With so many paths in each $I_{C}$ it is computationally
impractical to compare the image of every $S_{C}$ from all the sprinklings to
$S_{\mathcal{M}}$. To circumvent this problem we conduct our study using a
Monte Carlo style sampling of the paths in each trial $I_{C}$.\newpage

\begin{center}
Table 1. Average number of longest paths $S_{C}$ versus interval

size for ten trial sprinklings. The notation $a$ $(b)$ means $a\times10^{b}%
$.\medskip%

\begin{tabular}
[c]{cc}\hline\hline
$N$ & $\left\langle S_{C}\right\rangle $\\\hline
100 & \qquad6.70 (1)\\
500 & \qquad5.18 (3)\\
1000 & \qquad3.69 (6)\\
1500 & \qquad1.15 (8)\\
2000 & \qquad9.97 (9)\\\hline\hline
\end{tabular}
\bigskip
\end{center}

Our procedure is the following: (a) we obtain a trial path $s_{1}$ by
selecting the first elements (smallest label) in the various $Y_{Dj}$. (b)
Using the coordinates from which the elements of $s_{1}$ were sprinkled, we
calculate the goodness-of-fit of the image of $s_{1}$ to $S_{\mathcal{M}}$
using a $\chi^{2}$ statistic%
\begin{equation}
\chi^{2}=\frac{1}{n(d-1)}\sum_{j=1}^{n}\sum_{k=1}^{d-1}\frac{(x_{j}^{k}%
-y_{j}^{k})^{2}}{y_{j}^{k}} \label{chi2}%
\end{equation}
where $n$ is the number of elements in the longest path, $d-1$ is the number
of spatial dimensions of the manifold, the $x_{j}^{k}$ are the spatial
coordinates of points in the image of the path $s_{1}$ in $\mathcal{M}$ and
the $y_{j}^{k}$ are the spatial coordinates of those points along the geodesic
having the same time coordinate as the point with $x_{j}^{k}$. So, in standard
chi-squared language, the points from the path play the role of the
\textquotedblleft observed\textquotedblright\ values, the points from the
geodesic play the role of the \textquotedblleft expected\textquotedblright%
\ values, and $n$ is the number of \textquotedblleft degrees of
freedom.\textquotedblright\ (c) We then find a new path by randomly choosing
any element, from among all the elements that are along any longest path, that
has more than one youngest daughter, randomly choosing one of the youngest
daughters, then completing the path with the first elements of the remaining
$Y_{Dj}$. (d) We calculate the $\chi^{2}$ for this new path and save this path
if it is a better fit than the previous path, otherwise it is discarded. (e)
We repeat steps (c) and (d) for a number of trials equal to $5\times L(x_{1}%
)$. The prescription for the number of trials is adopted so that the number of
trials automatically scales upward as the number of elements in the causal
set, and therefore the number of $s_{j}$ to be sampled, increases; the
constant 5 is found to be sufficient to produce the best results. The path
that emerges from this process is taken as $S_{C}$ for that trial $I_{C}$. We
conduct multiple trial sprinklings (typically 100 for flat spacetimes and 1000
for curved) for a given size $N$ and compute the average of the $\chi^{2}$
values for each $S_{C}$. In flat spacetimes small enough that all of the
longest paths can be checked, this procedure sucessfully found the absolute
best fit in most of the cases and nearly so in the others \cite{greg}.

It is worth pointing out that there is a sense in which this approach attempts
to mimic the role of an eventual causal set dynamics. One expects that a
sum-over-causal sets dynamics should reveal a statistical preference for paths
near the geodesic in much the same way as the sum-over-histories approach in
quantum mechanics reveals a statistical preference for classical trajectories.
To gauge whether the expected\ longest path does approach the geodesic in the
large $N$ limit, we calculate the average chi-squared $<\chi^{2}>$ of all of
the paths sampled during the Monte Carlo process for all trial sprinklings of
a given $N$. What we expect is that, if the conjectures hold, the expected
longest path will show consistently better fits as $N$ increases. It is
important to keep in mind that it is the \textit{trend} of the fits that
matters here moreso than the fits themselves.

In flat spacetimes, it is straightforward to identify the actual geodesic
curves $S_{\mathcal{M}}$. For curved spacetimes the $S_{\mathcal{M}}$ are
found numerically using the shooting method. The analytical details of how the
geodesic equations are set up for doing these calculations are given elsewhere
\cite{raluca}. The specific coordinates $y_{j}^{k}$ used to calculate the
$\chi^{2}$ values in Eq. (\ref{chi2}) are then found by interpolating between
the numerically calculated points.

\section{Results and Conclusions}

\subsection{Flat Spacetimes}

Figure 1 shows the results for 1+1-dimensional Minkowski space.
Panel (a) shows a uniform sprinkling of 5000 points into an
interval together with the geodesic $S_{\mathcal{M}}$ and a best
longest path $S_{C}$ determined from the Monte Carlo process
described previously. We can see from this plot that, modulo
statistical fluctuations, $S_{C}$ produces a reasonably good
visual fit to $S_{\mathcal{M}}$. However, it is the decreasing
trend of the $<\chi^{2}>$ in panel (b) of this figure that clearly
establishes the key result that the expected longest path in the
interval approaches the geodesic as the number of causal set
elements increases. Similar results for 2+1- and 3+1-dimensional
Minkowski space are shown in Figure 2 and Figure 3, respectively.
For those cases (figures 2a, 3b, 3c, and 3d) only projections of
uniform sprinklings are shown. In those projections the points do
not visually appear to be as uniform as in figure 1a because in
the higher dimensional cases there is much more volume near the
center of the interval than near $x_{1}$ and $x_{N}$ causing more
of the projected points to be centrally located. As with the
1+1-dimensional case, we can clearly see in figures 2b and 3a that
$<\chi ^{2}>$ decreases as $N$ increases. These flat spacetime
calculations constitute direct numerical evidence in favor of the
conclusions reached by Brightwell and co-workers \cite{bng}.

\subsection{Curved Spacetimes}

Figure 4 shows the results for a 1+1-dimensional conformally flat
spacetime with conformal factor $\Omega^{2}=\left(  xt\right)
^{2}$. Panel (a) shows a uniform sprinkling of 5000 points.
Careful inspection of this sprinkling shows that most of the
volume (and therefore the points) occur at larger values of $x$
and $t$. Panel (b) shows the key result that the expected longest
path approaches the geodesic curve as $N$ increases. For a visual
sense of what occurs in this spacetime, panel (c) compares
$S_{\mathcal{M}}$ with $S_{C}$ also showing the corresponding flat
spacetime geodesic for comparison. This figure shows that this
$S_{C}$ does appear to follow $S_{\mathcal{M}}$, especially at
larger times where there are many more points in the sprinkling.

Figure 5 shows the results for a 2+1-dimensional conformally flat
spacetime with conformal factor $\Omega^{2}=\left(
x^{4}+y^{4}\right)  /t^{6}$. Panel (a) shows the projections of a
uniform sprinkling of 5000 points. These projections show that
most of the volume occurs at smaller times and larger values of
$x$ and $y$. Panel (b) shows the key result that the expected
longest path approaches the geodesic curve as $N$ increases. In
panels (c) and
(d) we see the $x-t$ and $y-t$ projections of $S_{\mathcal{M}}$ and an $S_{C}%
$. In a similar way as in the 1+1 case, we see that at smaller times, where
most of the points are, the best longest path does appear to follow along the
geodesic curve.

Figure 6 shows the results for a 3+1-dimensional conformally flat spacetime
with conformal factor $\Omega^{2}=t^{4}e^{x^{2}-8y}$. This conformal factor
increases with $x$ and $t$, decreases with $y$, and has no $z$ dependence.
Therefore, the features of the interval are that there is more volume at
larger values of $x$ and $t$, smaller values of $y$, and uniform in $z$.
Projections of a uniform sprinkling of 5000 points into an interval for this
spacetime, shown in panels (b), (c), and (d), reveal these features of the
interval. Panel (a) confirms that even for this spacetime, the expected
longest path approaches the geodesic curve as $N$ increases. This fact is
supported visually by panels (b), (c), and (d) which also show that our best
longest path is an excellent fit to the geodesic curve.

\subsection{Conclusions}

In this paper we have provided computational evidence that the path-geodesic
correspondence, implied by key ideas in causal set research, holds for flat
and conformally flat curved spacetimes. We further suggest that by
establishing that the longest maximal chains approach the actual geodesic
curves in the spacetimes we have considered, we have also provided evidence in
favor of the wider conjecture, the Hauptvermutung, at the core of causal set kinematics.

\begin{acknowledgments}
We thank Dr. Rafael D. Sorkin for providing an algorithm for
calculating the lengths of paths in a causal set and for further
motivation to work on problems of this nature. We are thankful for
the helpful suggestions of Natthi L. Sharma, Ernest Behringer,
James P. Sheerin, Daniel W. Kittell, and Cosmin Ilie. Partial
support from the U. S. National Science Foundation is gratefully
acknowledged.
\end{acknowledgments}

\section{Figure Captions}

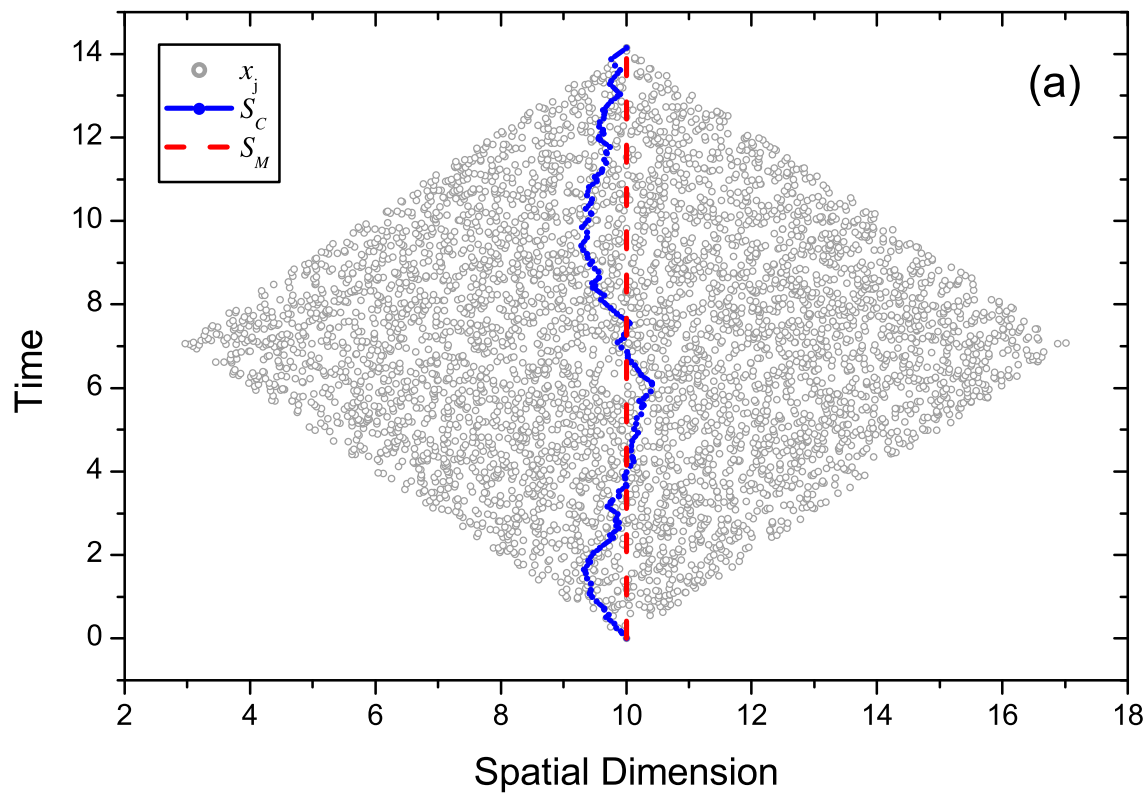
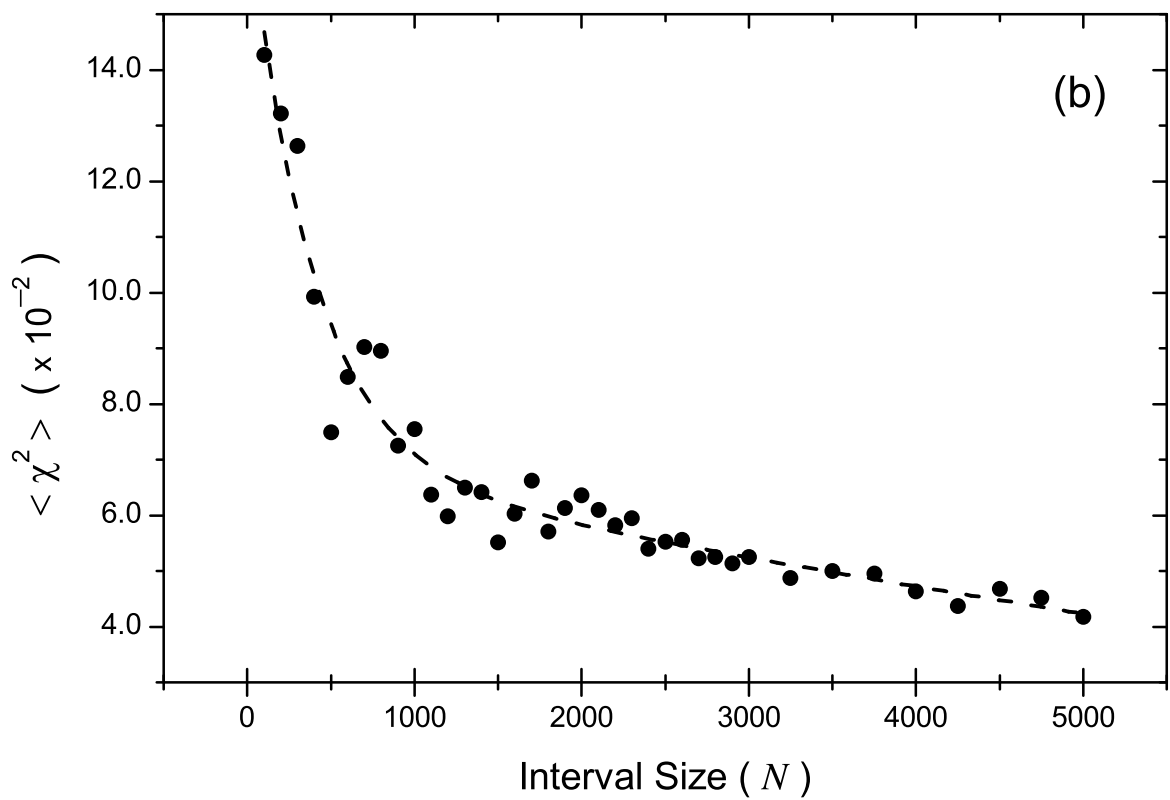
\textbf{FIG 1}. Results for 1+1-dimensional Minkowski space. Panel (a) shows
the sprinkled points $x_{j}$, the geodesic $S_{\mathcal{M}}$, and one of the
best longest paths $S_{C}$, determined by the Monte Carlo process described in
the text. Panel (b) shows the decrease in the $<\chi^{2}>$ versus interval
size $N$. The dashed line is just a numerical fit to the data points the
purpose of which is to emphasize the trend of the data.

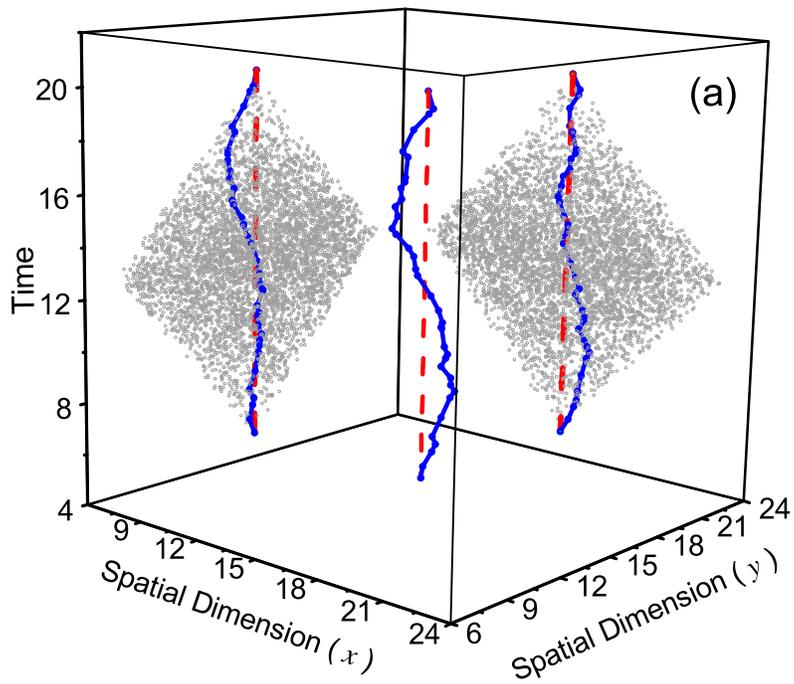
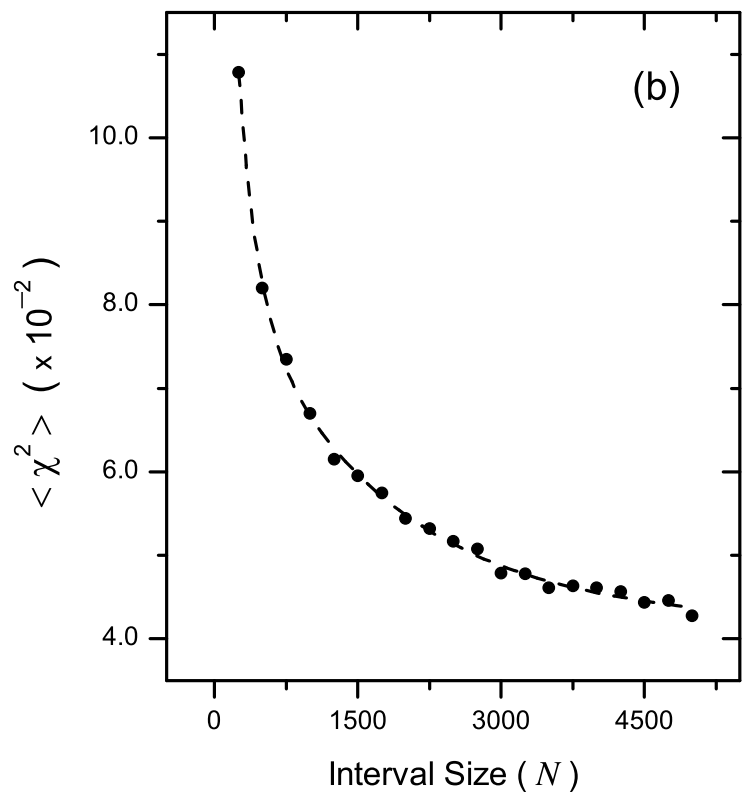
\textbf{FIG 2}. Results for 2+1-dimensional Minkowski space. Panel (a) shows
$x$-$t$ and $y$-$t$ projections of the sprinkled points, the geodesic and its
projections, and one of the best longest paths and its projections. Panel (b)
shows the decrease in the $<\chi^{2}>$ versus interval size $N$. The dashed
line emphasizes the trend of the data.

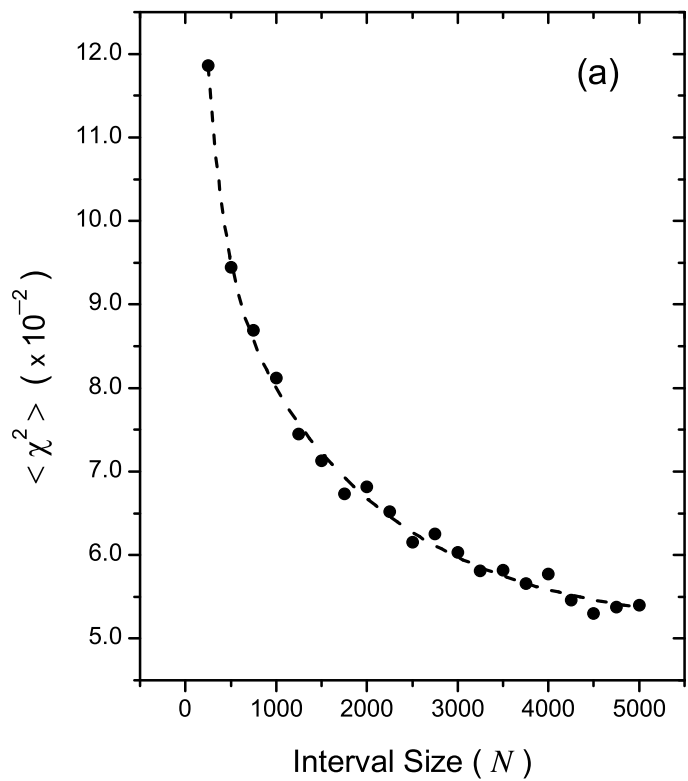
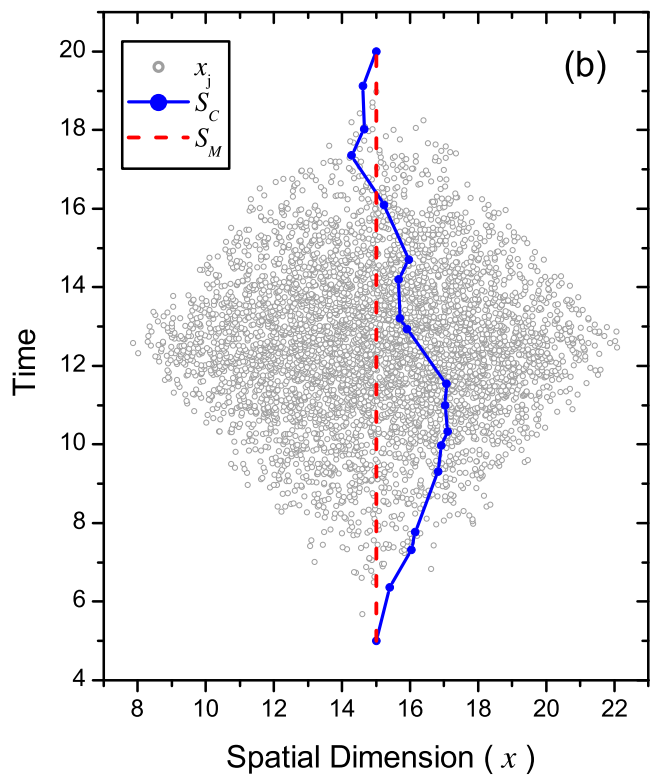
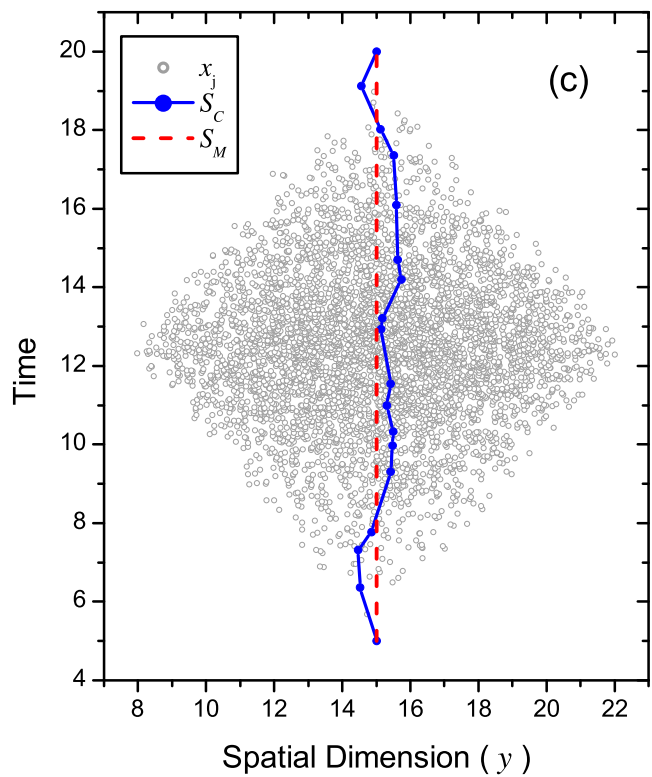
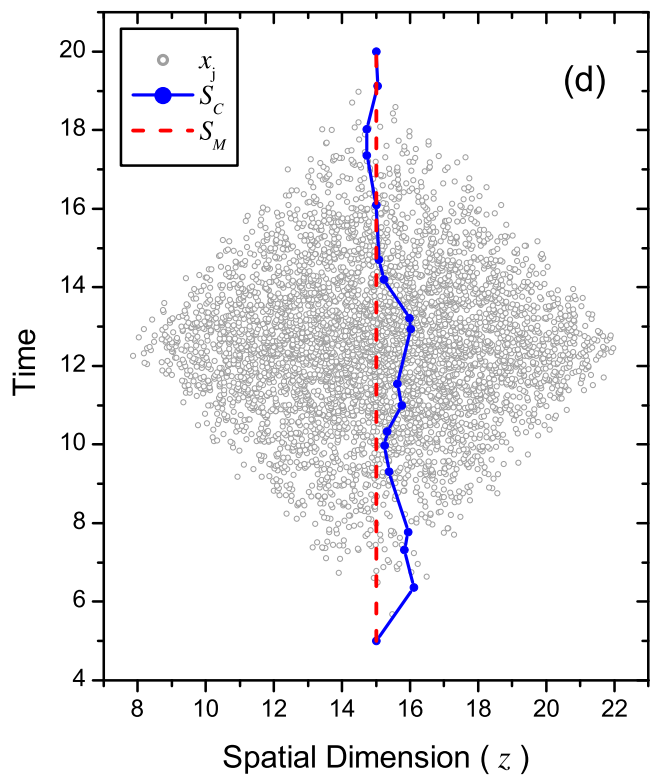
\textbf{FIG 3}. Results for 3+1-dimensional Minkowski space. Panel (a) shows
the decrease in the $<\chi^{2}>$ versus interval size $N$ with a dashed line
to emphasize the decreasing trend of the data. Panels (b) through (d) show
projections of the sprinkled points, the geodesic, and one of the best longest paths.

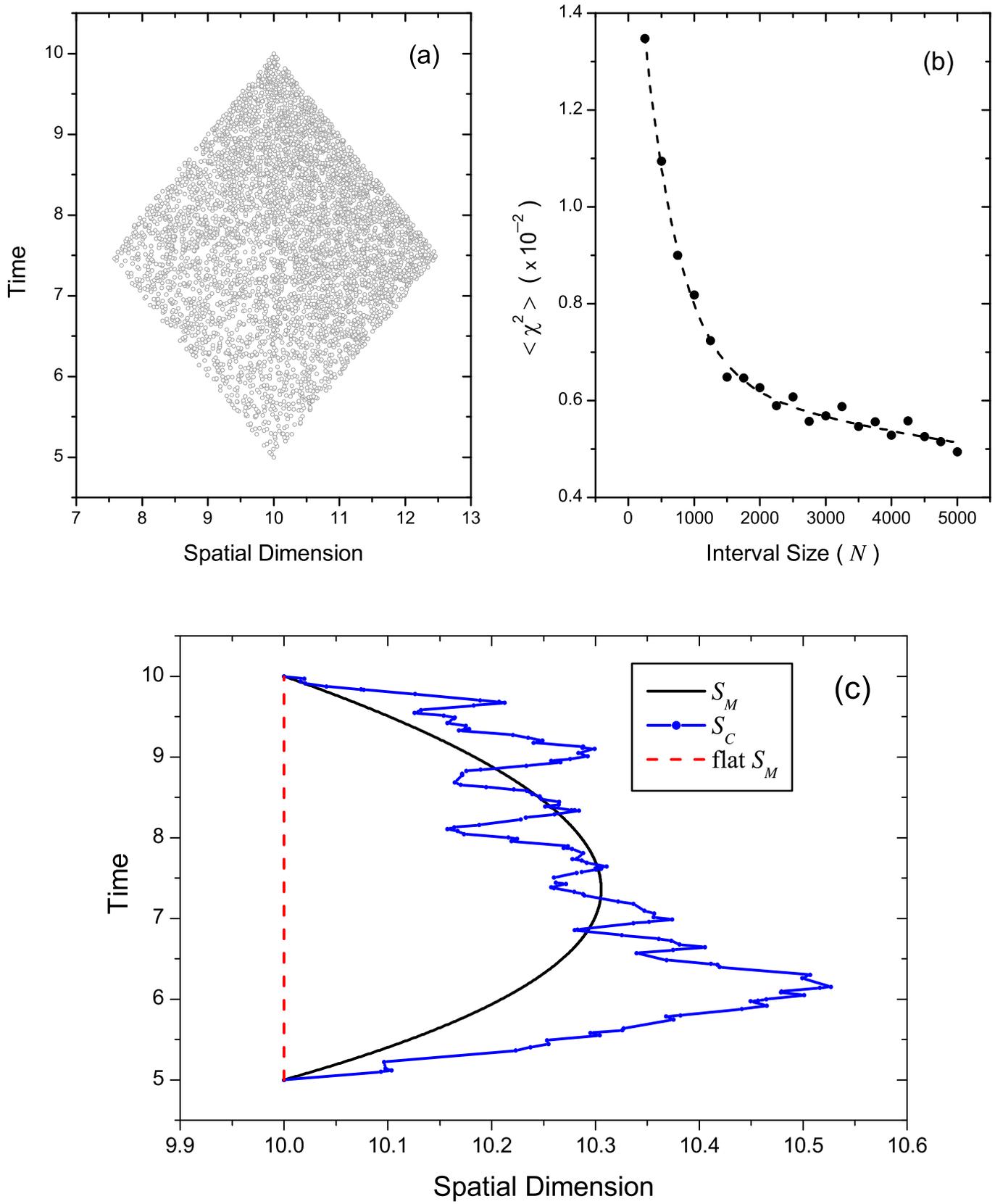
\textbf{FIG 4}. Results for 1+1-dimensional curved spacetime using the metric
of Eq. (\ref{conformal metric}) with conformal factor $\Omega^{2}=\left(
xt\right)  ^{2}$. Panel (a) shows the uniform sprinkling of 5000 points into
an interval. Panel (b) shows the decrease in the $<\chi^{2}>$ versus interval
size $N$. Panel (c) compares one of the best longest paths, $S_{C}$, to the
geodesics, $S_{\mathcal{M}}$, for both the curved spacetime and the
1+1-dimensional flat spacetime.

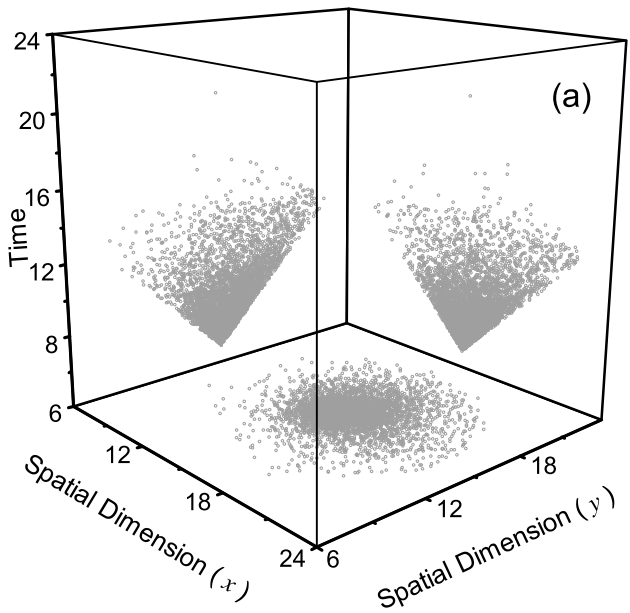
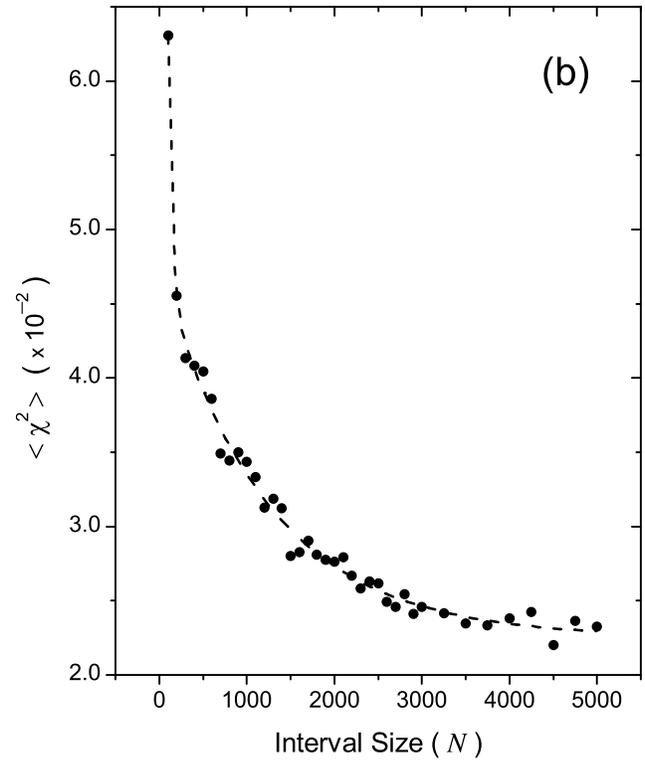
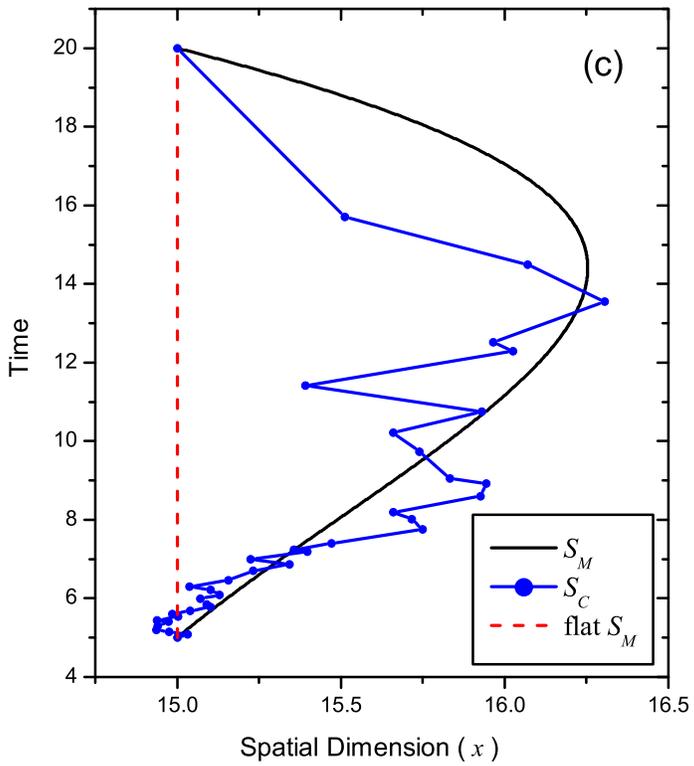
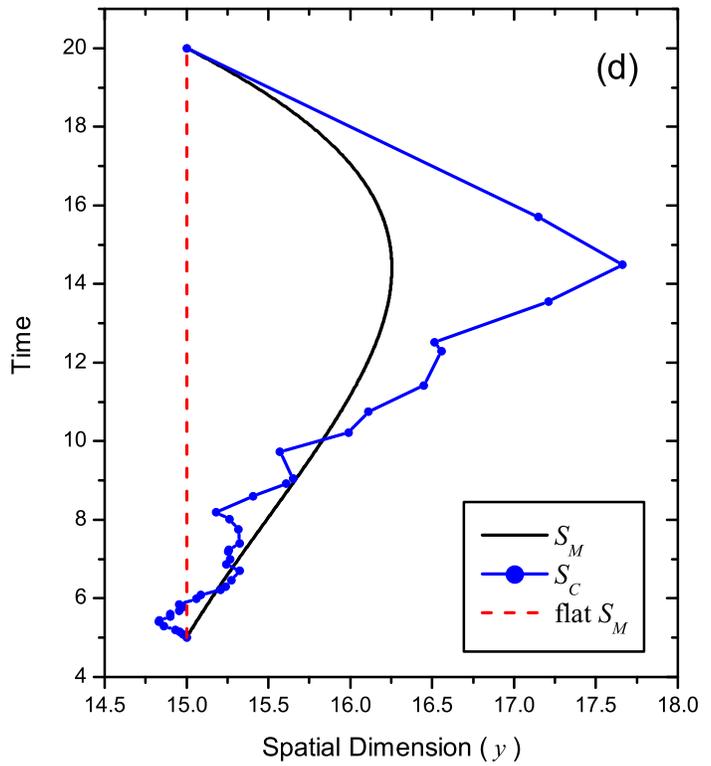
\textbf{FIG 5}. Results for 2+1-dimensional curved spacetime using the metric
of Eq. (\ref{conformal metric}) with conformal factor $\Omega^{2}=\left(
x^{4}+y^{4}\right)  /t^{6}$. Panel (a) shows $x$-$t$, $y$-$t$, and $x$-$y$
projections of a uniform sprinkling of 5000 points. Panel (b) shows the
decrease in the $<\chi^{2}>$ versus interval size $N$. Panels (c) and (d)
compare projections of one of the best longest paths, $S_{C}$, to those of the
geodesics, $S_{\mathcal{M}}$, for both the curved spacetime and the
2+1-dimensional flat spacetime.

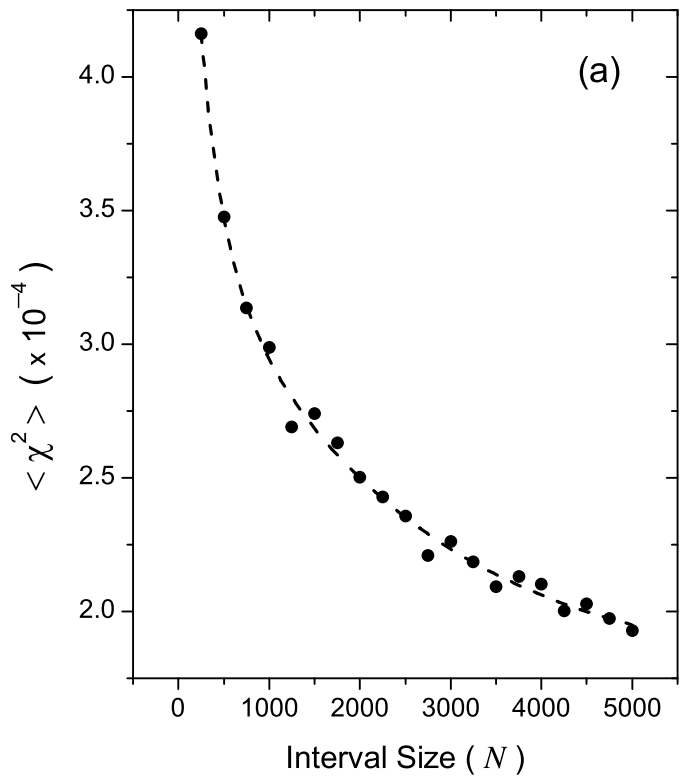
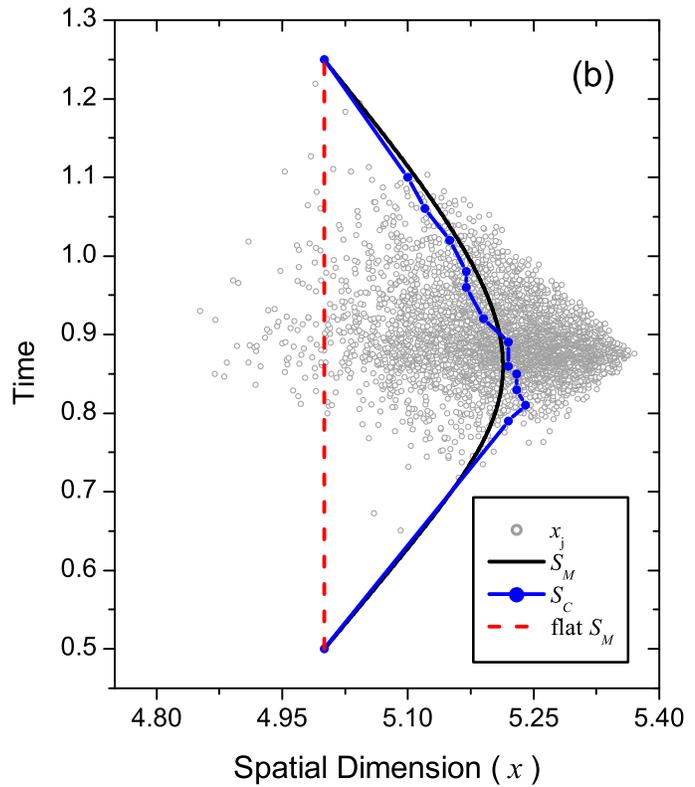
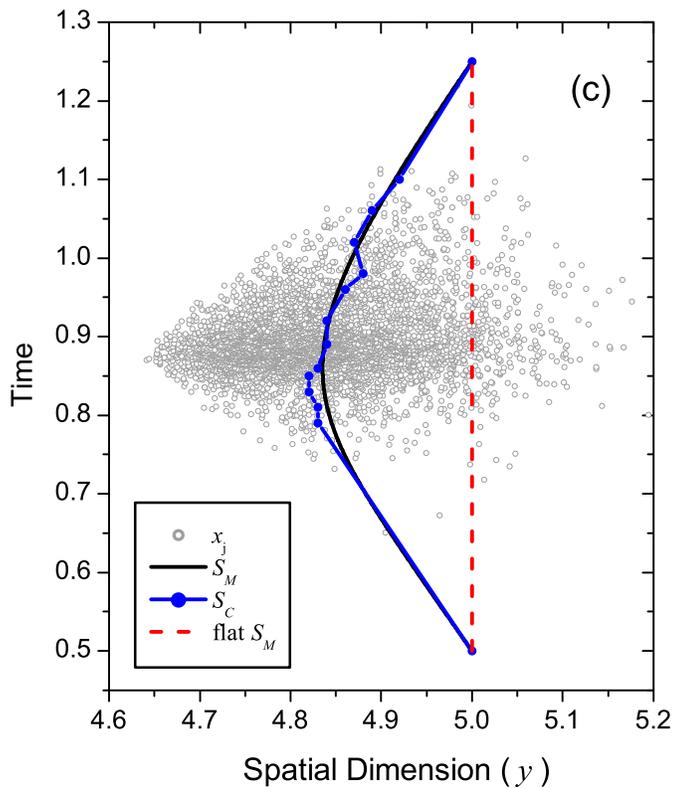
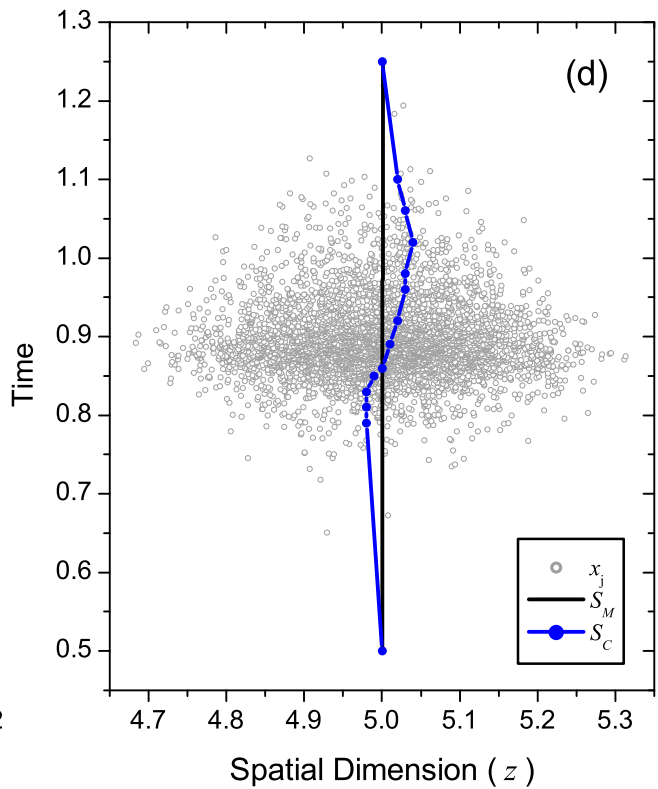
\textbf{FIG 6}. Results for 3+1-dimensional curved spacetime using the metric
of Eq. (\ref{conformal metric}) with conformal factor $\Omega^{2}%
=t^{4}e^{x^{2}-8y}$. Panel (a) shows the decrease in the $<\chi^{2}>$ versus
interval size $N$. Panels (b) through (d) show projections of the sprinkled
points and compares projections of one of the best longest paths, $S_{C}$, to
those of the geodesics, $S_{\mathcal{M}}$, for both the curved spacetime and
the 3+1-dimensional flat spacetime.

\end{document}